\documentclass[12pt]{article}
\usepackage{epsfig}

\renewcommand{\thesection} {\arabic{section}}

\def\beq{\begin{equation}}
\def\eeq{\end{equation}}
\def\bea{\begin{eqnarray}}
\def\eea{\end{eqnarray}}

\begin{document}

\begin{titlepage}

\begin{flushright}
WSU-HEP-0404\\
MCTP-04-36\\
{\tt hep-ph/0407004}\\
\end{flushright}

\vspace{1.5cm}
\begin{center}
\Large\bf\boldmath
Spectator effects and lifetimes of heavy hadrons
\unboldmath
\end{center}

\vspace{1.0cm}
\begin{center}
Fabrizio Gabbiani$^1$, Andrei I. Onishchenko$^{1}$ and Alexey A. Petrov$^{1,2}$\\[0.7cm]
{\sl $^1$Department of Physics and Astronomy, Wayne State University,\\
Detroit, MI 48201\\[0.1cm]

$^2$Michigan Center for Theoretical Physics, University of Michigan,\\
Ann Arbor, MI 48109}
\end{center}

\vspace{0.8cm}
\begin{abstract}
\vspace{0.2cm}
\noindent
We present a calculation of subleading $1/m_b$ corrections to spectator effects
in the ratios of beauty hadron lifetimes in heavy-quark expansion. We find that
these effects are sizable and should be taken into account in systematic analyses
of heavy hadron lifetimes. In particular, the inclusion of $1/m_b$ corrections
brings into agreement the theoretical predictions and experimental observations of
the ratio of lifetimes of $\Lambda_b$-baryon and $B_d$ meson. We
obtain $\tau(B_u)/\tau(B_d) = 1.06 \pm 0.02, \,
\tau(B_s)/\tau(B_d)= 1.00 \pm 0.01, \, \tau(\Lambda_b)/\tau(B_d)= 0.86
\pm 0.05$.
\end{abstract}

\end{titlepage}
\setcounter{page}{2} \newpage

\section{Introduction}

The hierarchy of lifetimes of heavy hadrons can be understood in
the heavy-quark expansion (HQE), which makes use of the disparity of scales present in the
decays of hadrons containing b-quarks. HQE predicts the ratios of lifetimes of beauty
mesons, which agree with the experimental observations well within experimental and
{\it theoretical} uncertainties. Most recent analyses give~\cite{PDG,ave}
\bea
\tau(B_u)/\tau(B_d)|_{ex}=1.085 \pm0.017,\quad && \tau(B_u)/\tau(B_d)|_{th}=1.07 \pm 0.03, \\
\tau(B_s)/\tau(B_d)|_{ex}=0.951 \pm 0.038,\quad  && \tau(B_s)/\tau(B_d)|_{th}=1.00 \pm 0.02,
\eea
which show evidence of excellent agreement of theoretical predictions and experimental
measurements. This agreement also provides us with some confidence that quark-hadron
duality, which states that smeared partonic amplitudes can be replaced by the hadronic
ones, is expected to hold in inclusive decays of heavy flavors. The heavy-quark expansion
also provides a qualitative understanding of the hierarchy of lifetimes of charmed
hadrons. Since the expansion parameter there is much larger, only qualitative agreement
is expected.

The only remaining problem appears in the ratios of lifetimes of heavy mesons
and baryons. For instance, the low experimental value of the ratio
$\tau(\Lambda_b)/\tau(B_d)$ has long been a puzzle for the theory. The latest
compilation of experimental observations~\cite{ave} suggests that
\beq
\tau(\Lambda_b)/\tau(B_d)|_{ex}= 0.797 \pm 0.053,
\eeq
which is inconsistent with the first theoretical predictions
of this ratio, $0.90-0.95$~\cite{Bigi:1994wa,Neubert:1996we,Rosner:1996fy} at the level of
approximately $2\sigma$. Only recent next-to-leading order (NLO) calculations of
perturbative QCD~\cite{Ciuchini:2001vx} and $1/m_b$ corrections~\cite{Gabbiani:2003pq}
to spectator effects significantly reduced the discrepancy, yielding, as reported in
\cite{Tarantino:2003qw},
\beq
\tau(\Lambda_b)/\tau(B_d)|_{th}= 0.88 \pm 0.05,
\eeq
which appears in some agreement with experimental data, mainly due to sizable theoretical
and experimental uncertainties. The problem of $\tau(\Lambda_b)/\tau(B_d)$ ratio
could reappear again if future measurements at Fermilab and CERN would find the mean
value to stay the same with error bars shrinking. Upcoming Fermilab measurements of
$\Lambda_b$ lifetime could shed more light on the experimental side of this issue.

Here we expand our calculation of subleading contributions to
spectator effects in the
$1/m_b$ expansion~\cite{Gabbiani:2003pq} to study their impact on the ratios
of lifetimes of heavy mesons. We include the full charm quark-mass dependence, generalizing
our result of Ref.~\cite{Gabbiani:2003pq}. We also discuss the convergence of the $1/m_b$
expansion in the analysis of spectator effects by estimating next, $1/m_b^2$-suppressed
corrections to the spectator effects in factorization.

This paper is organized as follows. We set up the relevant formalism and argue for
the need to compute subleading $1/m_b$ corrections to spectator effects in
Sect.~\ref{formalism}. In Sect.~\ref{AllResults} we discuss the impact of
$1/m_b$ corrections to lifetime ratios of heavy mesons and baryons and
assess the convergence of the $1/m_b$ expansion. We present the complete results from
$1/m_b^2$ corrections to spectator effects. Finally, we present our conclusions in
Sect.~\ref{Conclusions}. Some of the relevant formulas are shown in the Appendices.

\section{Formalism}\label{formalism}

The inclusive decay rate of a heavy hadron $H_b$ is most conveniently computed by employing
the optical theorem to relate the decay width to the imaginary part of the forward matrix
element of the transition operator:
\beq \label{rate}
\Gamma(H_b)=\frac{1}{2 M_{H_b}} \langle H_b |{\cal T} | H_b \rangle,~~
{\cal T} = {\mbox{Im}}~ i \int d^4 x T \left\{
H_{\mbox{\scriptsize eff}}(x) H_{\mbox{\scriptsize eff}}(0) \right \}.
\eeq
Here $H_{\mbox{\scriptsize eff}}$ represents an effective $\Delta B=1$ Hamiltonian,
\beq
H_{\mbox{\scriptsize eff}} = \frac{4 G_F}{\sqrt{2}} V_{cb} \sum_{d'=d,s, u'=u,c}
V^*_{u'd'} \left[ C_1(\mu) Q_1^{u'd'} (\mu) +
C_2(\mu) Q_2^{u'd'} (\mu) \right] + h.c.,
\eeq
where the four-quark operators $Q_1$ and $Q_2$ are given by
\beq
Q_1^{u'd'}=\bar d_{L}' \gamma_\mu u_{L}' ~\bar c_{L} \gamma^\mu b_{L},\qquad
Q_2^{u'd'}=\bar c_{L} \gamma_\mu u_{L}' ~\bar d_{L}' \gamma^\mu b_{L}.
\eeq
In the heavy-quark limit, the energy release is large, so the correlator in
Eq.~(\ref{rate}) is dominated by short-distance physics.
An Operator Product Expansion (OPE) can be constructed for Eq.~(\ref{rate}),
which results in a prediction of decay widths of Eq.~(\ref{rate}) as
a series of matrix elements of local operators of increasing dimension suppressed by powers
of $1/m_b$\footnote{We choose a heavy-quark pole mass as an expansion parameter,
which is consistent with most of the previous calculations. Of course, final
predictions for the lifetime ratios are independent of that choice~\cite{Bigi:1994em}.}:
\beq\label{expan}
\Gamma(H_b)= \frac{1}{2 M_{H_b}} \sum_k \langle H_b |{\cal T}_k | H_b \rangle
=\sum_{k} \frac{C_k(\mu)}{m_b^{k}}
\langle H_b |{\cal O}_k^{\Delta B=0}(\mu) | H_b \rangle.
\eeq
In other words, the calculation of $\Gamma(H_b)$ is equivalent to computing
the matching coefficients of the effective $\Delta B=0$ Lagrangian
with subsequent computation of its matrix elements. Indeed, at the end, the
scale dependence of the Wilson coefficients in Eq.~(\ref{expan}) should match
the scale dependence of the computed matrix elements.

It is customary to make predictions for the ratios of lifetimes (widths),
as many theoretical uncertainties cancel out in the ratio. In addition, since
the differences of lifetimes should come from the differences in the ``brown mucks''
of heavy hadrons, at the leading order in HQE all beauty hadrons with light
spectators have the same lifetime.

The difference between meson and baryon lifetimes first occurs at order
$1/m^2$ and is essentially due to the different structure of mesons and
baryons. In other words, no lifetime difference is induced among the members
of the meson multiplet. The ratio of heavy meson and baryon lifetimes
receives a shift which amounts to at most $1-2\%$, not sufficient
to explain the observed pattern of lifetimes.

The main effect appears at the $1/m^3$ level and comes from the set of
dimension-six four-quark operators. Their contribution is
enhanced due to the phase-space factor $16 \pi^2$, and induces
corrections of order $16 \pi^2 (\Lambda_{QCD}/m_b)^3$ = ${\cal O}(5-10\%)$.
These operators introduce a difference in lifetimes for both heavy mesons and baryons.
The effects of matrix elements of these operators, commonly referred to as
Weak Annihilation (WA), Weak Scattering (WS), and Pauli Interference (PI)
have been computed
in~\cite{Neubert:1996we,Guberina:1979xw,Bilic:1984nq,Guberina:1986gd,Shifman:wx,Beneke:2002rj,Guberina:2000de}
at leading order in perturbative QCD, and, more recently, including NLO perturbative
QCD corrections in~\cite{Ciuchini:2001vx} and $1/m_b$ corrections~\cite{Gabbiani:2003pq}.

The contributions of these operators to the lifetime ratios are
governed by the matrix elements of $\Delta B=0$ four-fermion operators
\bea
{\cal T}_{\rm spec} &=&
{\cal T}_{\rm spec}^{u} +
{\cal T}_{\rm spec}^{d'} +
{\cal T}_{\rm spec}^{s'},
\eea
where the ${\cal T}_{i}$ are
\bea\label{SpecLO}
{\cal T}_{\rm spec}^{u} &=&
\frac{G_F^2m_b^2|V_{bc}|^2(1-z)^2}{2\pi}
\Big\{
\left(c_1^2+c_2^2\right)
O_1^{u}
+2c_1c_2\widetilde{O}_1^{u}
+ \delta_{1/m}^{u}
+ \delta_{1/m^2}^{u}
\Big\}, \nonumber \\
{\cal T}_{\rm spec}^{d'} &=&
-\frac{G_F^2m_b^2|V_{bc}|^2(1-z)^2}{4\pi}
\left\{
c_1^2\left[
(1+z)O_1^{d'}+\frac{2}{3}(1+2z)O_2^{d'}
\right] \right. \nonumber \\ && \left.
+\left(
N_cc_2^2+2c_1c_2
\right)
\left[
(1+z)\widetilde{O}_1^{d'}+\frac{2}{3}(1+2z)\widetilde{O}_2^{d'}
\right]
+\delta_{1/m}^{d'} + \delta_{1/m^2}^{d'}
\right\}, \\
{\cal T}_{\rm spec}^{s'} &=&
-\frac{G_F^2m_b^2|V_{bc}|^2\sqrt{1-4z}}{4\pi}
\left\{
c_1^2\left[
O_1^{s'}+\frac{2}{3}(1+2z)O_2^{s'}
\right] \right. \nonumber \\ && \left.
+\left(N_cc_2^2+2c_1c_2
\right)
\left[
\widetilde{O}_1^{s'}+\frac{2}{3}(1+2z)\widetilde{O}_2^{s'}
\right]
+\delta_{1/m}^{s'} + \delta_{1/m^2}^{s'}
\right\}. \nonumber
\eea
Here the terms $\delta^i_{1/m}$ and $\delta^i_{1/m^2}$ refer to $1/m_b$ and $1/m_b^2$
corrections to spectator effects, which we discuss below. Note that we include
the full $z = {m_c^2}/{m_b^2}$ dependence, which is fully consistent only after the
inclusion of higher $1/m_b$ corrections. The operators $O_i$ and $\widetilde{O}_i$ in
Eq.~(\ref{SpecLO}) are defined as
\bea
O_1^q = \bar b_i\gamma^{\mu}(1-\gamma_5)b_i\bar
q_j\gamma_{\mu}(1-\gamma_5)q_j, \quad &&
O_2^q = \bar b_i\gamma^{\mu}\gamma_5b_i\bar
q_j\gamma_{\mu}(1-\gamma_5)q_j,\nonumber \\
\widetilde{O}_1^q = \bar b_i\gamma^{\mu}(1-\gamma_5)b_j
\bar q_i\gamma_{\mu}(1-\gamma_5)q_j, \quad &&
\widetilde{O}_2^q = \bar b_i\gamma^{\mu}\gamma_5 b_j
\bar q_i\gamma_{\mu}(1-\gamma_5)q_j.
\eea
In addition, there is a dimension-6 operator
$O_3^q = \bar b_i\gamma^{\mu}(1-\gamma_5)b_i \bar q_j\gamma_{\mu}\gamma_5q_j$ whose
contribution is always suppressed by powers of the light-quark mass. It gives
a negligible contribution to the ratios of lifetimes of heavy hadrons
(see also Sect.~\ref{AllResults}).

In order to assess the impact of these and other operators, parametrizations of
their matrix elements must be introduced. The meson matrix elements are
\bea \label{MEM}
\langle B_q |O_1^q| B_q \rangle &=& f^2_{B_q} m^2_{B_q} \left(2 \epsilon_1 + \frac{B_1}{N_c}
\right), ~~
\langle B_q |\widetilde{O}_1^q| B_q \rangle = f^2_{B_q} m^2_{B_q} B_1,
\nonumber \\
\langle B_q |O_2^q| B_q \rangle &=& -f^2_{B_q} m^2_{B_q} \left[
\frac{m_{B_q}^2}{\left(m_b+m_q\right)^2} \left(2 \epsilon_2 + \frac{B_2}{N_c}\right)
+ \frac{1}{2}\left(2 \epsilon_1 + \frac{B_1}{N_c}
\right)
\right], \\
\langle B_q |\widetilde{O}_2^q| B_q \rangle &=& -f^2_{B_q} m^2_{B_q} \left[
\frac{m_{B_q}^2}{\left(m_b+m_q\right)^2} B_2 + \frac{1}{2} B_1
\right]. \nonumber
\eea
Here the parameters $B_i$ and $\epsilon_i$ are usually referred to as
``singlet'' and ``octet'' parameters (see~\ref{fqope} for the explanation).
Expressed in terms of these parameters, the lifetime ratios of heavy mesons
can be written as
\bea \label{MesonMeson}
\tau(B_u)/\tau(B_d)= 1+16 \pi^2 \frac{f_B^2 m_B}{m_b^3 c_3(m_b)}
\left[G_1^{ss} (m_b) B_1 (m_b)+G_1^{oo} (m_b) \epsilon_1 (m_b)
\right.
\nonumber \\
\left. + ~G_2^{ss} (m_b) B_2 (m_b)+ G_1^{oo} (m_b) \epsilon_2 (m_b) \right] + \delta_{1/m},
\eea
where the coefficients $G$ were computed at NLO
in~\cite{Beneke:2002rj,Ciuchini:2001vx}.
$\delta_{1/m}$ represents spectator corrections of order $1/m_b$ and higher,
which we present below in this paper.

Estimates of the matrix elements of four-fermion operators in baryon decays are
not easy. Following~\cite{Neubert:1996we}, the parameter $\widetilde{B}$ is used
to account for the deviation of the $\Lambda_b$ wave function from being
totally color-asymmetric,
\bea
\langle \Lambda_b |O_1^q| \Lambda_b \rangle &=&
-\widetilde{B} \langle \Lambda_b |\widetilde{O}_1^q| \Lambda_b \rangle =
\frac{\widetilde{B}}{6}  f^2_{B_q} m_{B_q} m_{\Lambda_b} r, \quad
\nonumber \\
\langle \Lambda_b |O_2^q| \Lambda_b \rangle &=&
-\widetilde{B} \langle \Lambda_b |\widetilde{O}_2^q| \Lambda_b \rangle=
\frac{\widetilde{B}}{6} f^2_{B_q} m_{B_q} m_{\Lambda_b} \delta.
\eea
Note that $\delta={\cal O}(1/m_b)$, which follows from the heavy-quark spin
symmetry. It needs to be included as we consider higher-order corrections
in $1/m_b$. $\widetilde{B}=1$ in the valence approximation.
While these parameters have not been computed model-independently,
various quark-model arguments suggest that the meson and baryon matrix elements
are quite different. Thus a meson-baryon lifetime difference can be
produced. In general, one can parametrize the meson-baryon lifetime
ratio as
\bea \label{MesonBaryon}
\tau(\Lambda_b)/\tau(B_d) &\simeq&  0.98 - (d_1+d_2 \widetilde{B})r
-(d_3 \epsilon_1+d_4 \epsilon_2) -(d_5 B_1 + d_6 B_2) \nonumber \\
 &\simeq&  0.98 - m_b^2 (d_1'+d_2' \widetilde{B})r
-m_b^2 \left[(d_3' \epsilon_1+d_4' \epsilon_2) +(d_5' B_1 + d_6' B_2)\right],
\eea
where in the last line we scaled out the coefficient $m_b^2$ emphasizing
the fact that these corrections are suppressed by
$1/m_b^3$ compared to the leading $m_b^5$ effect. The scale-dependent
parameters ($d_i(m_b)$ =  \{0.023, 0.028, 0/16, --0.16, 0.08, --0.08\} at NLO)
are defined in~\cite{Neubert:1996we}. The parameter
$r=\left|\psi^{\Lambda_b}_{bq}(0)\right|^2/\left|\psi^{B_q}_{b\bar q}(0)\right|^2$
is the ratio of the wave functions at the origin of the $\Lambda_b$ and $B_q$ mesons.

It is interesting to note that in the absence of $1/m_b$ corrections to spectator
effects, it would be equally correct to substitute the $b$-quark mass in
Eq.~(\ref{MesonBaryon}) with the corresponding meson and baryon masses, so
\bea\label{MesonBaryon2}
\tau(\Lambda_b)/\tau(B_d) \simeq
0.98 - m_{\Lambda_b}^2 (d_1'+d_2' \widetilde{B})r
-m_{B_d}^2 \left[(d_3' \epsilon_1+d_4' \epsilon_2) +(d_5' B_1 + d_6' B_2)\right],
\eea
which reflects the fact that WS and PI effects occur for the heavy and light quarks
initially bound in the $B_d$ meson and $\Lambda_b$ baryon, respectively. While correct
up to the order $1/m_b^3$, these simple substitutions reduce the ratio of lifetimes
by approximately 3-4\%! We take this as an indication of the importance of bound-state
effects on the spectator corrections, represented by subleading $1/m_b$ corrections
to spectator operators.

\section{Subleading corrections to spectator effects}\label{AllResults}

We computed the higher order corrections, including charm quark-mass effects,
to Eq.~(\ref{SpecLO}) in the heavy-quark expansion, denoted below as
$\delta_{1/m}^{q}$ and $\delta_{1/m^2}^{q}$.

\subsection{\mbox{\boldmath $1/m_b$} corrections}

\begin{figure}[tb]
\centerline{\epsfxsize=9cm\epsffile{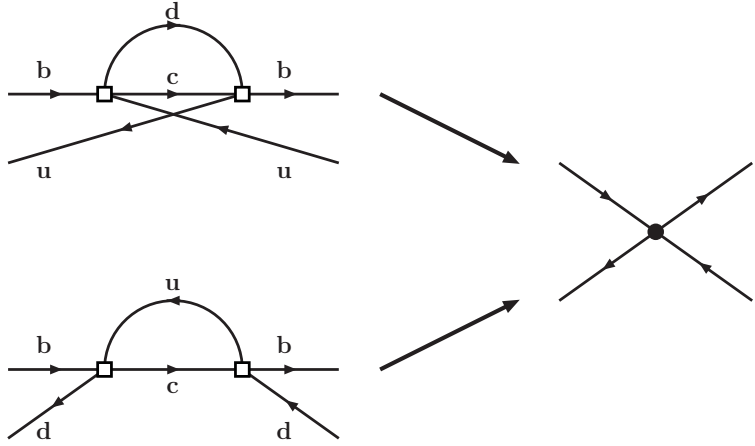}}
\centerline{\parbox{14cm}{\caption{\label{fig:WA}
Kinetic corrections to spectator effects. The operators of Eqs.~(\ref{OurCorrection})
and (\ref{OurCorrection1}) are obtained by expanding the diagrams in powers of
spectator's momentum.}}}
\end{figure}

The $1/m_b$ corrections to the spectator effects are computed, as in Ref.~\cite{Gabbiani:2003pq},
by expanding the forward scattering amplitude of Eq.~(\ref{rate}) in the light-quark momentum
and matching the result onto the operators containing derivative insertions
(see Fig.~\ref{fig:WA}). The $\delta_{1/m}^{q}$ contributions can be written in
the following form:
\bea\label{OurCorrection}
\delta_{1/m}^{u} &=&
-2\left(c_1^2+c_2^2\right)\frac{1+z}{1-z}R_1^{u}
-4c_1c_2\frac{1+z}{1-z}\widetilde{R}_1^{u}, \nonumber \\
\delta_{1/m}^{d'} &=&
c_1^2\left[
\frac{8z^2}{1-z}R_0^{d'}
+\frac{2}{3}\frac{1+z+10z^2}{1-z}R_1^{d'}
+\frac{2}{3}(1+2z)\left(
R_2^{d'}-R_3^{d'}
\right)
\right], \nonumber \\ &&
+\left(N_cc_2^2+2c_1c_2\right)
\left[
\frac{8z^2}{1-z}\widetilde{R}_0^{d'}
+\frac{2}{3}\frac{1+z+10z^2}{1-z}\widetilde{R}_1^{d'}
+\frac{2}{3}(1+2z)\left(
\widetilde{R}_2^{d'}-\widetilde{R}_3^{d'}
\right)
\right], \\
\delta_{1/m}^{s'} &=&
c_1^2\left[
\frac{16z^2}{1-4z}R_0^{s'}
+\frac{2}{3}\frac{1-2z+16z^2}{1-4z}R_1^{s'}
+\frac{2}{3}(1+2z)
\left(
R_2^{s'}-R_3^{s'}
\right)
\right] \nonumber \\ &&
+\left(N_cc_2^2+2c_1c_2\right)
\left[
\frac{16z^2}{1-4z}\widetilde{R}_0^{s'}
+\frac{2}{3}\frac{1-2z+16z^2}{1-4z}\widetilde{R}_1^{s'}
+\frac{2}{3}(1+2z)
\left(
\widetilde{R}_2^{s'}-\widetilde{R}_3^{s'}
\right)
\right],\nonumber
\eea
where the following operators contribute
\bea
R_0^q = \frac{1}{m_b^2}\bar b_i\gamma^{\mu}\gamma_5\vec{D}^{\alpha}b_i
\bar q_j\gamma_{\mu}(1-\gamma_5)\vec{D}_{\alpha}q_j, ~~~~~~~ \quad &&
R_1^q = \frac{1}{m_b^2}\bar b_i\gamma^{\mu}(1-\gamma_5)\vec{D}^{\alpha}b_i
\bar q_j\gamma_{\mu}(1-\gamma_5)\vec{D}_{\alpha}q_j, \nonumber \\
R_2^q = \frac{1}{m_b^2}\bar b_i\gamma^{\mu}(1-\gamma_5)\vec{D}^{\nu}b_i
\bar q_j\gamma_{\nu}(1-\gamma_5)\vec{D}_{\mu}q_j, \quad &&
R_3^q = \frac{m_q}{m_b}\bar b_i(1-\gamma_5)b_i\bar q_j(1-\gamma_5)q_j.
\eea
Here $\widetilde{R}_i^q$ denote the color-rearranged
operators that follow from the expressions for $R_i^q$ by
interchanging color indexes of $b_i$ and $q_j$ Dirac spinors.
Note that the above result contains {\it full} QCD $b$-fields, thus there
is no immediate power counting available for these operators. The power
counting becomes manifest at the level of the matrix elements.

In the $z\to 0$ limit the result of Eq.~(\ref{OurCorrection}) reduces to the
set of $1/m_b$ corrections obtained by us in Ref.~\cite{Gabbiani:2003pq}.
We complete the calculation presented there by computing the contribution of
Eq.~(\ref{OurCorrection}) to the ratios of heavy-meson lifetimes.
As before, we use factorization to guide our parametrizations of $\Lambda_b$
and meson matrix elements, but, contrary to Ref.~\cite{Gabbiani:2003pq}, we
elect to keep corrections that would vanish in the vacuum-insertion approximation.
Our parametrizations for meson and baryon matrix elements are presented in
\ref{matelf}. Inserting Eqs.~(\ref{MEMeson}) and (\ref{MEBaryon})
of \ref{matelf} into Eq.~(\ref{OurCorrection}) we obtain an estimate
of the effect. We postpone the presentation of the answer, complete with our
estimate of its uncertainties, until the end of this section, after we assess the
convergence of the $1/m_b$ expansion for the spectator corrections.

Numerically, the set of $1/m_b$ corrections does not markedly affect the ratios of
meson lifetimes, changing the $\tau(B_u)/\tau(B_d)$ and the $\tau(B_s)/\tau(B_d)$
ratios by less than half a percent. The effect is more pronounced in
the ratio of $\Lambda_b$ and $B_d$ lifetimes, where it constitutes a $40-45\%$ of the
leading spectator contribution represented by WS and PI effects
(see Eqs.~(\ref{MesonBaryon},\ref{MesonBaryon2})), or an overall correction of about $-(3-4)\%$
to the $\tau(\Lambda_b)/\tau(B_d)$ ratio. While such a sizable effect is
surprising, the main source of such a large correction can be readily identified.
While the individual $1/m_b$ corrections to the matrix elements representing
WS and PI are of order $20\%$, as expected from the naive power counting, they
contribute to the $\Lambda_b$ lifetime with the same (negative) sign, instead of
destructively interfering as do the contributions to Eqs.~(\ref{MesonBaryon},\ref{MesonBaryon2})
representing WS and PI~\cite{Rosner:1996fy}, as explained in~\cite{Gabbiani:2003pq}.
This conspiracy of two $\sim 20\%$ effects produces a sizable shift in the ratio of the
$\Lambda_b$ and $B$-meson lifetimes.

\subsection{\mbox{\boldmath $1/m^2_b$} corrections}

In order to see how well the $1/m_b$ expansion converges in the calculation of lifetime
ratios, we compute a set of $\delta_{1/m_b^2}$ corrections to spectator effects.
It is expected that at this order even more operators will contribute,
so the exact prediction of lifetime differences becomes unfeasible.
Therefore, we shall use our calculation only to see to what extent we
can trust our results for the $1/m_b$ corrections. We will parametrize
$1/m_b^2$ corrections similar to our parametrization of $1/m_b$
effects above and will use the factorization approximation to assess their
contributions to the lifetime ratios.

There are two classes of corrections that arise at this order. One class involves kinetic
corrections which can be computed in a way analogous to the previous case by expanding the
forward scattering amplitudes in the powers of spectator momentum (see Fig.~\ref{fig:WA}).
A second class involves corrections arising from an interaction with background gluon
fields. The kinetic corrections can be written as:
\bea\label{OurCorrection1}
\delta_{1/m^2}^{u} &=&
\left(c_1^2+c_2^2\right)
\left[
\frac{m_u^2}{m_b^2}\frac{1+z}{1-z}O_1^{u}
+\frac{8z^2}{(1-z)^2}W_1^{u}
\right]
+2c_1c_2
\left[
\frac{m_u^2}{m_b^2}\frac{1+z}{1-z}\widetilde{O}_1^{u}
+\frac{8z^2}{(1-z)^2}\widetilde{W}_1^{u}
\right],  \nonumber \\
\delta_{1/m^2}^{d'} &=&
c_1^2\left[
\frac{m_{d'}^2}{m_b^2}\frac{1+z+2z^2}{1-z}O_1^{d'}
+\frac{m_{d'}^2}{m_b^2}\frac{4z^2}{1-z}O_2^{d'}
+\frac{m_{d'}^2}{m_b^2}\frac{2(1+2z)}{3}O_3^{d'}
\right. \nonumber \\ && \left.
+\frac{4z^2(7z-5)}{(1-z)^2}W_1^{d'}
+\frac{8z^2(4z-3)}{(1-z)^2}W_2^{d'}
+\frac{8z^2}{1-z}\left(W_3^{d'}-W_4^{d'}\right)
\right] \nonumber \\ &&
+\left(N_cc_2^2+c_1c_2\right)
\left[
\frac{m_{d'}^2}{m_b^2}\frac{1+z+2z^2}{1-z}\widetilde{O}_1^{d'}
+\frac{m_{d'}^2}{m_b^2}\frac{4z^2}{1-z}\widetilde{O}_2^{d'}
+\frac{m_{d'}^2}{m_b^2}\frac{2(1+2z)}{3}\widetilde{O}_3^{d'}
\right. \nonumber \\ && \left.
+\frac{4z^2(7z-5)}{(1-z)^2}\widetilde{W}_1^{d'}
+\frac{8z^2(4z-3)}{(1-z)^2}\widetilde{W}_2^{d'}
+\frac{8z^2}{1-z}\left(\widetilde{W}_3^{d'}-\widetilde{W}_4^{d'}\right)
\right],  \nonumber \\
\delta_{1/m^2}^{s'} &=&
c_1\left[
\frac{m_{s'}^2}{m_b^2}\frac{1-2z}{1-4z}O_1^{s'}
+\frac{m_{s'}^2}{m_b^2}\frac{8z^2}{1-4z}O_2^{s'}
+\frac{m_{s'}^2}{m_b^2}\frac{2(1+2z)}{3}O_3^{s'}
\right. \nonumber \\ && \left.
+\frac{8z^2(16z-5)}{(1-4z)^2}W_1^{s'}
+\frac{16z^2(10z-3)}{(1-4z)^2}W_2^{s'}
+\frac{16z^2}{1-4z}\left(W_3^{s'}-W_4^{s'}\right)
\right] \nonumber \\ &&
+\left(N_cc_2^2+2c_1c_2\right)
\left[
\frac{m_{s'}^2}{m_b^2}\frac{1-2z}{1-4z}\widetilde{O}_1^{s'}
+\frac{m_{s'}^2}{m_b^2}\frac{8z^2}{1-4z}\widetilde{O}_2^{s'}
+\frac{m_{s'}^2}{m_b^2}\frac{2(1+2z)}{3}\widetilde{O}_3^{s'}
\right. \nonumber \\ && \left.
+\frac{8z^2(16z-5)}{(1-4z)^2}\widetilde{W}_1^{s'}
+\frac{16z^2(10z-3)}{(1-4z)^2}\widetilde{W}_2^{s'}
+\frac{16z^2}{1-4z}\left(\widetilde{W}_3^{s'}-\widetilde{W}_4^{s'}\right)
\right].
\eea
We again retain the dependence on quark masses in the above expression, including the
terms proportional to light-quark masses (and thus multiplied by the operators $O_i$).
The operators in Eq.~(\ref{OurCorrection1}) are defined as
\bea
W_1^q = \frac{1}{m_b^4}
\bar b_i\gamma^{\mu}(1-\gamma_5)\vec{D}^{\alpha}\vec{D}^{\beta}b_i
\bar q_j\gamma_{\mu}(1-\gamma_5)\vec{D}_{\alpha}\vec{D}_{\beta}q_j, &&
W_2^q = \frac{1}{m_b^4}
\bar b_i\gamma^{\mu}\gamma_5\vec{D}^{\alpha}\vec{D}^{\beta}b_i
\bar q_j\gamma_{\mu}(1-\gamma_5)\vec{D}_{\alpha}\vec{D}_{\beta}q_j,
\nonumber \\
W_3^q = \frac{1}{m_b^4}\bar b_i\gamma^{\mu}(1-\gamma_5)
\vec{D}^{\nu}\vec{D}^{\alpha}b_i
\bar q_j\gamma_{\nu}(1-\gamma_5)\vec{D}_{\mu}\vec{D}_{\alpha}q_j, &&
W_4^q = \frac{m_q}{m_b}\bar b_i(1-\gamma_5)\vec{D}^{\alpha}b_i
\bar q_j(1-\gamma_5)\vec{D}_{\alpha}q_j,
\eea
where, as before, $\widetilde{W}_i^q$ denote the color-rearranged operators that follow
from the expressions for $W_i^q$ by interchanging the color indexes of $b_i$ and $q_j$
Dirac spinors. The parametrization of matrix elements of these operators can be found
in~\ref{matelf}.

\begin{figure}[tb]
\centerline{\epsfxsize=9cm\epsffile{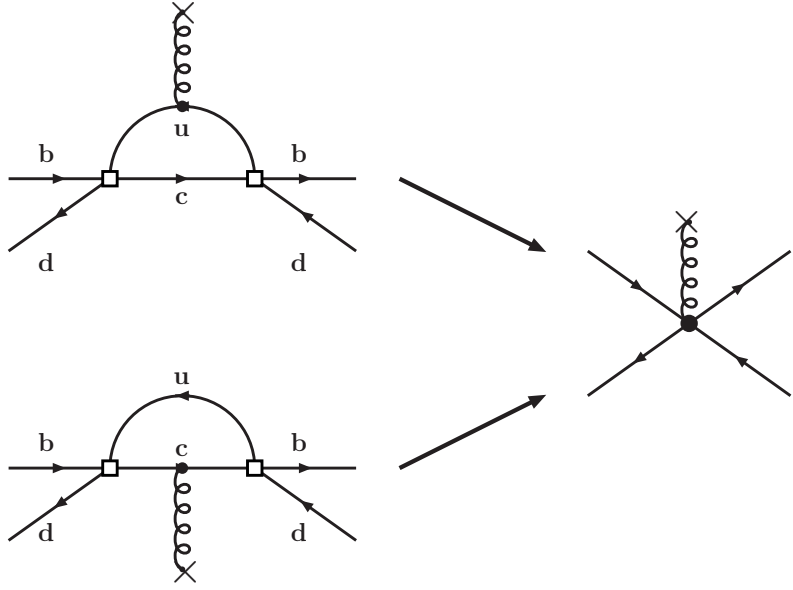}}
\centerline{\parbox{14cm}{\caption{\label{fig:WA2}
Corrections to spectator contribution coming from the interactions with
background gluon fields.}}}
\end{figure}
In addition to the set of kinetic corrections considered above, the effects of the
interactions of the intermediate quarks with background gluon fields should also
be included at this order. The contribution of those operators can be computed
from the diagrams of Fig.~\ref{fig:WA2}, resulting in
\bea
{\cal T}_{{\rm spec}, G} =
{\cal T}_{{\rm spec}, G}^u
+{\cal T}_{{\rm spec}, G}^{d'}
+{\cal T}_{{\rm spec}, G}^{s'}\;,
\eea
where
\bea
{\cal T}_{{\rm spec}, G}^u &=& 0, \\
{\cal T}_{{\rm spec}, G}^{d'} &=&
-\frac{G_F^2|V_{bc}|^2}{4\pi}\left\{
c_1^2\left[
(1-z^2)P_1^{d'}
-(1-z^2)P_2^{d'}
+2z(1-z)P_3^{d'}
+4z^2P_4^{d'}
\right] \right. \nonumber \\ && \left.
+2c_1c_2z\left[
(1-z)P_5^{d'}+(1-z)P_6^{d'}
+2zP_7^{d'}+2zP_8^{d'}
\right]
\right\}, \\
{\cal T}_{{\rm spec}, G}^{s'} &=&
-\frac{G_F^2|V_{bc}|^2}{4\pi\sqrt{1-4z}}
\left\{
c_1^2\left[
(1-2z)P_1^{s'}
-(1-4z)P_2^{s'}
+2zP_3^{s'}
+4z P_4^{s'}
\right] \right. \nonumber \\ && \left.
+4c_1c_2z\left[
P_7^{s'}+P_8^{s'}+P_9^{s'}+P_{10}^{s'}
\right]
\right\}.
\eea
The local four-quark operators in the above formulas
are shown in Eq.~(\ref{poper}).
\bea \label{poper}
P_1^q &=& \bar b_k\gamma^{\mu}(1-\gamma_5)t_{kl}^a\widetilde{G}_{\mu\nu}^a~b^{\phantom{l}}_l~
\bar q^{\phantom{l}}_i\gamma^{\nu}(1-\gamma_5)q^{\phantom{l}}_i, \nonumber\\
P_2^q &=& \bar b_i\gamma^{\mu}(1-\gamma_5)b^{\phantom{l}}_i~
\bar q^{\phantom{l}}_k\gamma^{\nu}(1-\gamma_5)t_{kl}^a\widetilde{G}_{\mu\nu}^a~q^{\phantom{l}}_l,
\nonumber \\
P_3^q &=& \bar b_k\gamma^{\mu}(1+\gamma_5)t_{kl}^a\widetilde{G}_{\mu\nu}^a~b_l~
\bar q^{\phantom{l}}_i\gamma^{\nu}(1-\gamma_5)q^{\phantom{l}}_i,  \nonumber\\
P_4^q &=& \frac{1}{m_b^2}\bar b_i\gamma^{\mu}(1-\gamma_5)
\vec{D}^{\nu}\vec{D}^{\alpha}b^{\phantom{l}}_i
\bar q^{\phantom{l}}_k\gamma_{\alpha}(1-\gamma_5)t_{kl}^a\widetilde{G}_{\mu\nu}^a q^{\phantom{l}}_l,
\nonumber\\
P_5^q &=& \bar b_k\gamma^{\mu}(1+\gamma_5)b^{\phantom{l}}_i~
t_{kl}^a\widetilde{G}_{\mu\nu}^a~
\bar q^{\phantom{l}}_i\gamma^{\nu}(1-\gamma_5)q^{\phantom{l}}_l, \nonumber\\
P_6^q &=& \bar b_i\gamma^{\mu}(1+\gamma_5)b_k~
t_{kl}^a\widetilde{G}_{\mu\nu}^a~
\bar q^{\phantom{l}}_l\gamma^{\nu}(1-\gamma_5)q^{\phantom{l}}_i,  \nonumber\\
P_7^q &=& \frac{1}{m_b^2}\bar b_k\gamma^{\mu}(1-\gamma_5)
\vec{D}^{\nu}\vec{D}^{\alpha}b^{\phantom{l}}_i~
t_{kl}^a\widetilde{G}_{\mu\nu}^a~
\bar q^{\phantom{l}}_i\gamma_{\alpha}(1-\gamma_5)q^{\phantom{l}}_l, \nonumber \\
P_8^q &=& \frac{1}{m_b^2}\bar b_i\gamma^{\mu}(1-\gamma_5)
\vec{D}^{\nu}\vec{D}^{\alpha}b^{\phantom{l}}_k~
t_{kl}^a\widetilde{G}_{\mu\nu}^a~
\bar q^{\phantom{l}}_l\gamma_{\alpha}(1-\gamma_5)q_i, \phantom{l} \nonumber\\
P_9^q &=& \bar b_k\gamma^{\mu}b_i~t_{kl}^a\widetilde{G}_{\mu\nu}^a~
\bar q^{\phantom{l}}_i\gamma^{\nu}(1-\gamma_5)q^{\phantom{l}}_l, \nonumber\\
P_{10}^q &=& \bar b_i\gamma^{\mu}b_k~t_{kl}^aG_{\mu\nu}^a
\bar q^{\phantom{l}}_l\gamma^{\nu}(1-\gamma_5)q^{\phantom{l}}_i.
\eea
Analogously to the previous section, we parametrize the matrix
elements as
\bea
\langle B_q | P_i^{d'} | B_q \rangle = \frac{1}{4} f^2_{B_{q}} m^2_{B_{q}}
\alpha^{q}_i \left(\frac{m_{B_q}^2}{m^2_b}-1\right)^2,~
\langle \Lambda_b | P_i^{q} | \Lambda_b \rangle = \frac{1}{4} f^2_{B_{q}}
m_{B_{q}} m_{\Lambda_b} \alpha^{\Lambda}_i
\left(\frac{m_{\Lambda_b}^2}{m^2_b}-1\right)^2.
\eea
We set $\alpha^{q}_i=\alpha^{\Lambda}_i = 1$~GeV$^2$ to obtain a numerical estimate
of this effect.
It is clear that no precise prediction is possible with so many operators
contributing to the ratios of lifetimes. This, of course, is expected, as
the number of contributing operators always increase significantly with each
order in OPE. Fortunately, at least in factorization,
the effect of $1/m_b^2$ corrections is tiny, a fraction
of a percent. The contribution of $1/m_b^2$-suppressed effect
is included into the coefficient $k_0$ defined in the next section.
We can evaluate the contribution of both $1/m_b$ and $1/m_b^2$ by performing two analyses.
First, we can study the perturbative scale dependence of the final result using
``factorized'' values for the matrix elements\footnote{In the absence of model-independent
calculations of the matrix elements described above, there is a number of ways to estimate them.
We chose to use the quark model to relate the matrix elements of $R_i$ and $\widetilde{R}_i$
operators to the matrix elements of dimension 6 operators $O_i$, which are parametrized
in terms of $B_i$ and $\epsilon_i$. We then use lattice QCD predictions for $B_i$ and
$\epsilon_i$ to obtain numerical estimates of our corrections. We shall refer to this
procedure as ``factorization.''}. We shall see that the scale dependence of
the resulting effect is very mild. Second, we shall randomly vary the parameters describing
the matrix elements by $\pm 30\%$ around their ``factorized'' values to see the uncertainty
of our result.

\subsection{Phenomenology of heavy hadron lifetimes}

Let us now discuss the phenomenological implications of the results presented in
the previous sections. As usual in the OPE-based calculation, next-order corrections
bring new unknown coefficients. In our numerical results we assume the value of
the $b$-quark pole mass to be $m_b=4.8\pm0.1$~GeV and $f_B=200\pm25$~MeV.

The resulting lifetime ratios are parametrized according to
\ref{matelf}. For instance, in the case of $\tau(\Lambda_b)/\tau(B_d)$ we can write
\bea\label{rLBd}
\left. \frac{\tau(\Lambda_b)}{\tau(B_d)}\right|_{h.o.} = k_0 +
\sum_{i=1} k_i p_i\ ,
\eea
where $p_i$ are the parameters defined in \ref{matelf},
\bea
p_i=\left\{
B_1, B_2, \epsilon_1, \epsilon_2, r, \delta,
\widetilde{B} r, \widetilde{B} \delta, \beta_{0}, \widetilde{\beta}_{0},
\beta_{1}, \widetilde{\beta}_{1}, \beta_{2}, \widetilde{\beta}_{2}, \beta_{3},
\widetilde{\beta}_{3}, \beta^{\Lambda}_{0}, \widetilde{\beta}^{\Lambda}_{0},
\beta^{\Lambda}_{1}, \widetilde{\beta}^{\Lambda}_{1},
\beta^{\Lambda}_{2}, \widetilde{\beta}^{\Lambda}_{2},
\beta^{\Lambda}_{3}, \widetilde{\beta}^{\Lambda}_{3}
\right\},
\eea
and $k_i$ are their coefficients. $k_0 = 1.00$ in the case of mesons, while for
$\Lambda_b$ $k_0 = \{0.991,0.981,0.975\}$ for $\mu = \{m_b/2,m_b, 2\,m_b\}$, respectively.
They appear in Table~\ref{rLBdT}, and
a similar procedure is followed for $\tau(B_u)/\tau(B_d)$ and
$\tau(B_s)/\tau(B_d)$, whose coefficients are shown in
Tables~\ref{rBudT} and \ref{rBsdT}, respectively.
\begin{table}[t]
\footnotesize
\begin{center}
\begin{tabular}{|c|cccccccccccc|}
\hline\hline
$~$ & \multicolumn{12}{|c|}{\mbox{Coefficients} $k_i$ \mbox{for the
parameters listed below. All entries are} $\times 10^{-3}$.} \\
$\mu$ & $B_1$ & $B_2$ & $\epsilon_1$ & $\epsilon_2$ & $r$ & $\delta$
& $\widetilde{B} r$ & $\widetilde{B} \delta$ & $\beta_{0}$ & $\widetilde{\beta}_{0}$ &
$\beta_{1}$ & $\widetilde{\beta}_{1}$ \\
\hline
$m_b/2$ & --7.82 & 9.19 & --158 & 183 & --27.4 & 3.76 &
--33.3 & 11.1 & 0.001 & 1.95 & --0.014 & --24.0 \\
$m_b$ & --7.96 & 9.31 & --160 & 196 & --22.4 &
3.39 & --27.8 & 12.4 & 0.031 & 1.78 & --0.382 & --21.9 \\
$2 m_b$ & --7.45 & 8.63 & --160 & 201 & --18.9 &
3.33 & --24.6 & 13.0 & 0.079 & 1.69 & --0.976 & --20.9 \\
\hline\hline
$\mu$ & $\beta_{2}$ & $\widetilde{\beta}_{2}$ & $\beta_{3}$ &
$\widetilde{\beta}_{3}$ &
$\beta^{\Lambda}_{0}$ & $\widetilde{\beta}^{\Lambda}_{0}$ &
$\beta^{\Lambda}_{1}$ & $\widetilde{\beta}^{\Lambda}_{1}$ &
$\beta^{\Lambda}_{2}$ & $\widetilde{\beta}^{\Lambda}_{2}$ &
$\beta^{\Lambda}_{3}$ & $\widetilde{\beta}^{\Lambda}_{3}$ \\
\hline
$m_b/2$ & --0.002 & --3.68 & 0.002 & 3.68 & --0.144 &
--0.048 & --12.7 & --6.85 & --0.818 & --0.272 & --0.818  & --0.272 \\
$m_b$ & --0.058 & --3.36 & 0.058 & 3.36 & --0.132 &
--0.039 & --11.2 & --4.59 & --0.747 & --0.223 & --0.747 & --0.223 \\
$2 m_b$ & --0.150 & --3.20 & 0.150 & 3.20 & --0.125
& --0.030 & --10.4 & --3.09 & --0.711 & --0.170 & --0.711 & --0.170 \\
\hline\hline

\end{tabular}
\end{center}
\caption{Coefficients $k_i$ appearing in Eq.~(\ref{rLBd}) for
$\tau(\Lambda_b)/\tau(B_d)$.}
\label{rLBdT}
\end{table}
\begin{table}[t]
\footnotesize
\begin{center}
\begin{tabular}{|c|cccccccccccc|}
\hline\hline
$~$ & \multicolumn{12}{|c|}{\mbox{Coefficients} $k_i$ \mbox{for the
parameters listed below. All entries are} $\times 10^{-3}$.} \\
$\mu$ & $B_1$ & $B_2$ & $\epsilon_1$ & $\epsilon_2$ & $\beta_{0}$ & $\widetilde{\beta}_{0}$ &
$\beta_{1}$ & $\widetilde{\beta}_{1}$ & $\beta_{2}$ & $\widetilde{\beta}_{2}$ & $\beta_{3}$ &
$\widetilde{\beta}_{3}$ \\
\hline
$m_b/2$ & 82.1 & 18.8 & --782 & 206 & 0.001 & 1.95 &
--17.6 & 124 & --0.002 & --3.68 & 0.002 & 3.68 \\
$m_b$ & 60.8 & 13.3 & --727 & 211 & 0.031 &
1.78 & --6.58 & 107 & --0.058 & --3.36 & 0.058 & 3.36 \\
$2 m_b$ & 46.5 & 9.98 & --693 & 213 & 0.079 &
1.69 & 0.661 & 98.9 & --0.150 & --3.20 & 0.150 & 3.20 \\
\hline\hline
\end{tabular}
\end{center}
\caption{Same as Table~\ref{rLBdT} for $\tau(B_u)/\tau(B_d)$.}
\label{rBudT}
\end{table}
\begin{table}[t]
\footnotesize
\begin{center}
\begin{tabular}{|c|cccccccl|}
\hline\hline
$~$ & \multicolumn{8}{|c|}{\mbox{Coefficients} $k_i$ \mbox{for the
parameters listed below. All entries are} $\times 10^{-3}$.} \\
$\mu$ & $B_1$ & $B_2$ & $\epsilon_1$ & $\epsilon_2$ & $B^s_1$ &
$B^s_2$ & $\epsilon^s_1$ & $\epsilon^s_2$ \\
\hline
$m_b/2$ & --7.82 & 9.19 & --158 & 184 & 11.2 & --12.4 &
180 & --230 \\
$m_b$ & --7.96 & 9.31 & --160 & 196 & 13.5 &
--16.8 & 182 & --243 \\
$2 m_b$ & --7.45 & 8.63 & --160 & 201 & 14.7 &
--19.1  & 183 & --249 \\
\hline\hline
$\mu$ & $\beta_{0}$ & $\widetilde{\beta}_{0}$ &
$\beta_{1}$ & $\widetilde{\beta}_{1}$ & $\beta_{2}$ & $\widetilde{\beta}_{2}$ & $\beta_{3}$ &
$\widetilde{\beta}_{3}$ \\
\hline
$m_b/2$ & --0.003 & --5.72 & 0.011 & 18.9 & 0.0013 & 2.20 &
--0.0013 & --2.20 \\
$m_b$ & --0.091 & --5.23 & 0.300 & 17.3 & 0.0349 &
2.00 & --0.0349 & --2.00 \\
$2 m_b$ & --0.233 & --4.97 & 0.768 & 16.4 & 0.0892 &
1.91 & --0.0892 & --1.91 \\
\hline\hline
\end{tabular}
\end{center}
\caption{Same as Table~\ref{rLBdT} for $\tau(B_s)/\tau(B_d)$.}
\label{rBsdT}
\end{table}
Since all three heavy mesons belong to the same $SU(3)$ triplet, their lifetimes
are the same at order $1/m_b^2$. The computation of the ratios of heavy meson
lifetimes is equivalent to the computation of $U$-spin or isospin-violating
corrections. Both $1/m_b^3$-suppressed spectator effects and our
corrections computed in the previous sections arise from the spectator interactions
and thus provide a source of $U$-spin or isospin-symmetry breaking. We shall, however,
assume that the matrix elements of both $1/m_b^3$ and $1/m_b^4$ operators respect
isospin.
The ratio of lifetimes of $B_s$ and $B_d$ mesons involves a breaking of $U$-spin symmetry,
so the matrix elements of dimension-6 operators could differ by about $30\%$. We partially
take this breaking into account by retaining light quark-mass dependence. Nevertheless,
we still introduce different $B-$ and $\epsilon$-parameters to describe $B_s$ and
$B_d$ lifetimes. We will also assume that, apart from the explicit light quark-mass
effects, matrix elements of $1/m_b^4$ operators respect $U$-spin symmetry.

\begin{figure}[!htb]
\centerline{\epsfxsize=14cm\epsffile{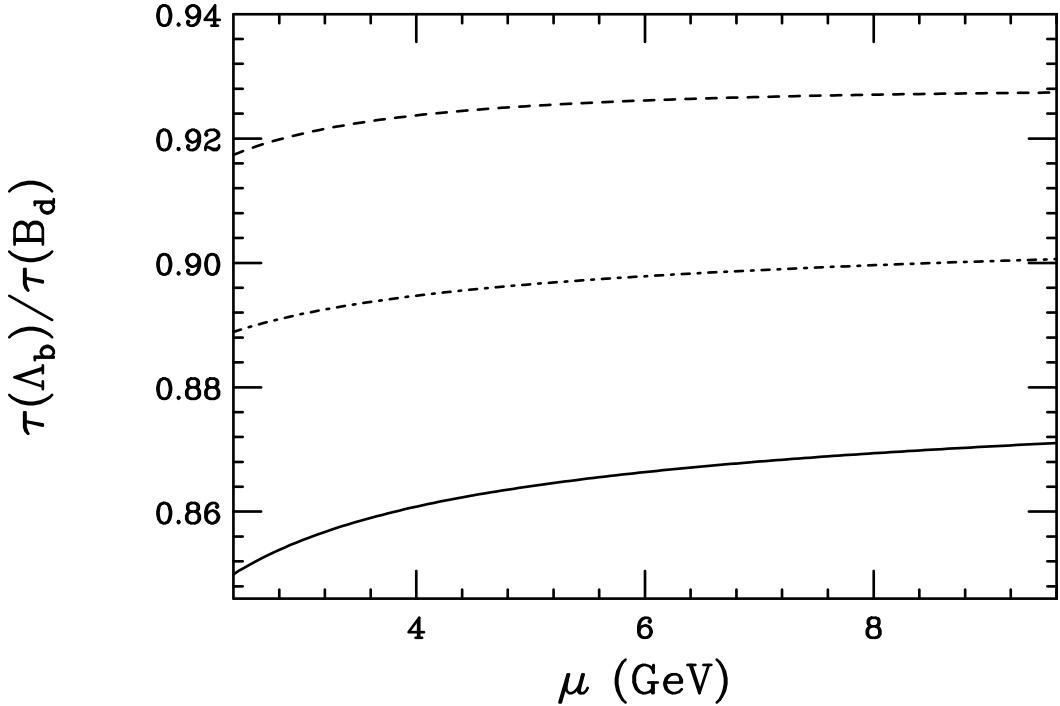}}
\centerline{\parbox{14cm}{\caption{\label{fig:mu}
LL (dashed line), NLL (dash-dotted line), and NLL+$1/m_b$
(solid line) contributions to $\tau(\Lambda_b)/\tau(B_d)$ are
plotted vs. the scale parameter $\mu$ in GeV.}}}
\end{figure}

\begin{figure}[!htb]
\centerline{\epsfxsize=14cm\epsffile{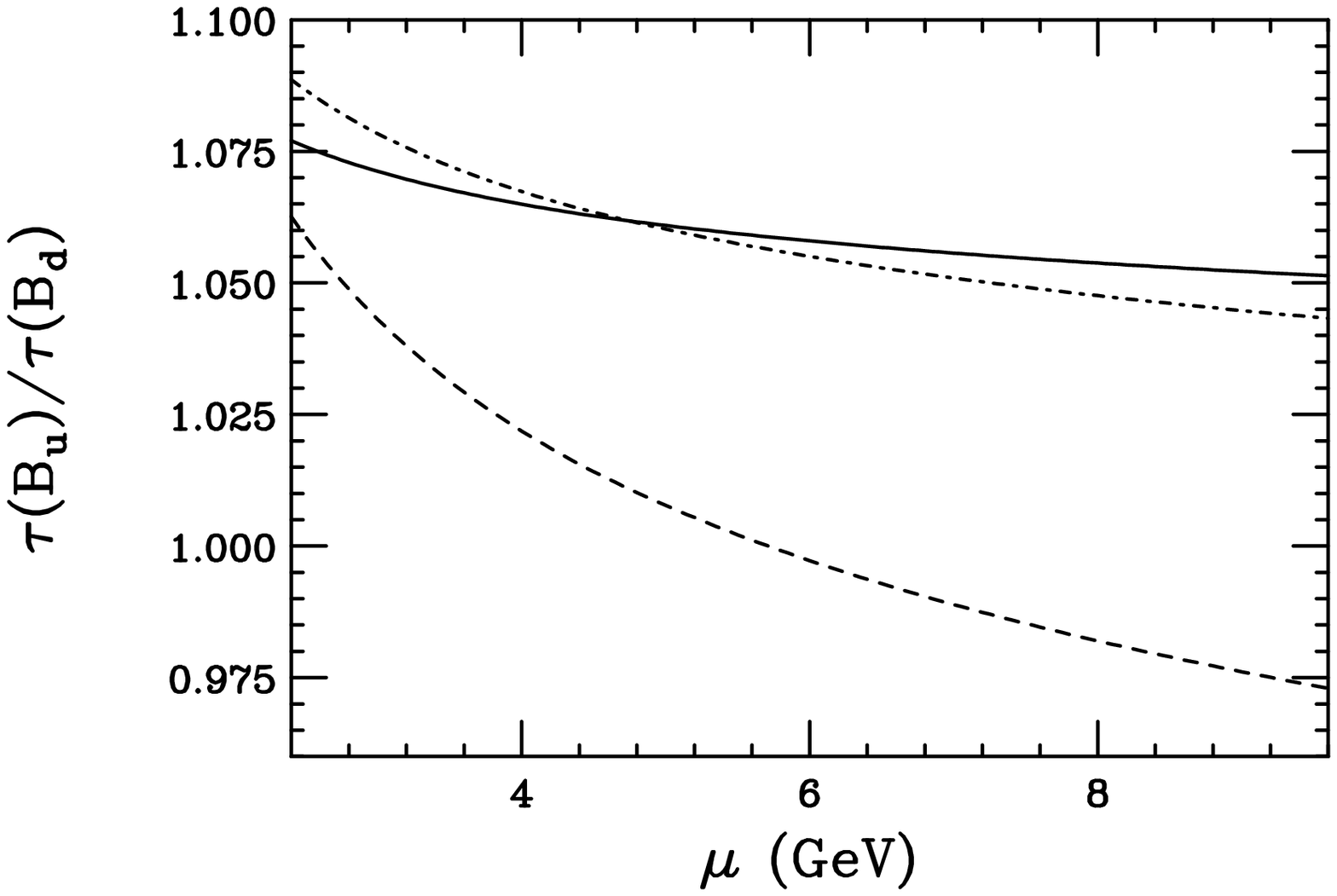}}
\centerline{\parbox{14cm}{\caption{\label{fig:mup}
Same as Fig.~\ref{fig:mu} for $\tau(B_u)/\tau(B_d)$.}}}
\end{figure}

\begin{figure}[!htb]
\centerline{\epsfxsize=14cm\epsffile{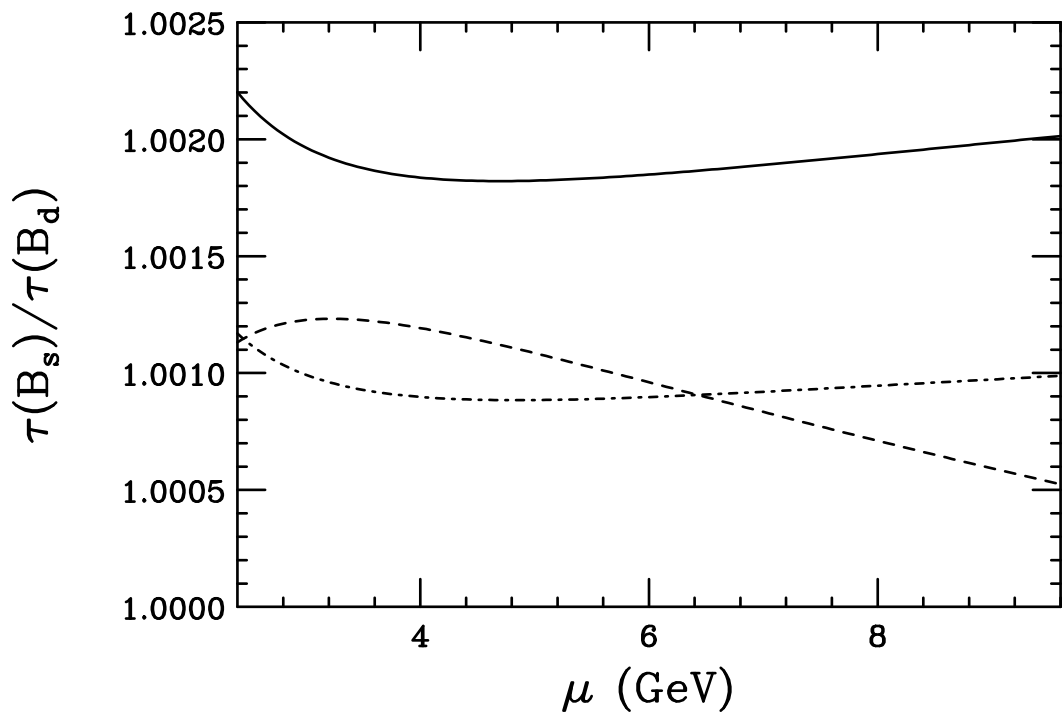}}
\centerline{\parbox{14cm}{\caption{\label{fig:mus}
Same as Fig.~\ref{fig:mu} for $\tau(B_s)/\tau(B_d)$.}}}
\end{figure}

\begin{figure}[!htb]
\centerline{\epsfig{file=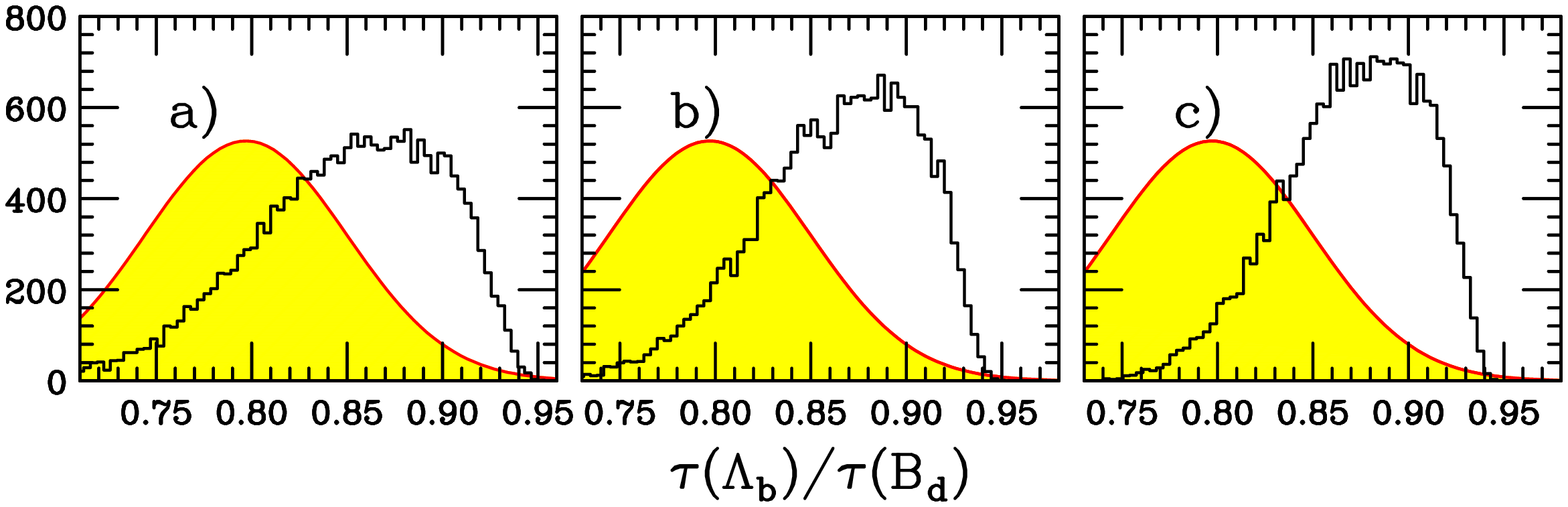, angle=90, width=16cm}}
\centerline{\parbox{14cm}{\caption{\label{fig:randLam}
Histograms showing the random distributions around the central values
of the $f_{B_q}$, $m_b$, $B$, $\delta$, $\epsilon$$, \beta_i$
and $\widetilde{\beta}_i$ parameters of
Eqs.~(\ref{MEMeson}) and (\ref{MEBaryon}) contributing to $\tau(\Lambda_b)/\tau(B_d)$.
Three histograms are shown for the scales $\mu = m_b/2$ (a), $\mu =
m_b$ (b), and $\mu = 2\, m_b$ (c). The shaded curves, presented for the convenience,
represent current experimental result. Their normalization is arbitrary.
}}}
\end{figure}

\begin{figure}[!htb]
\centerline{\epsfig{file=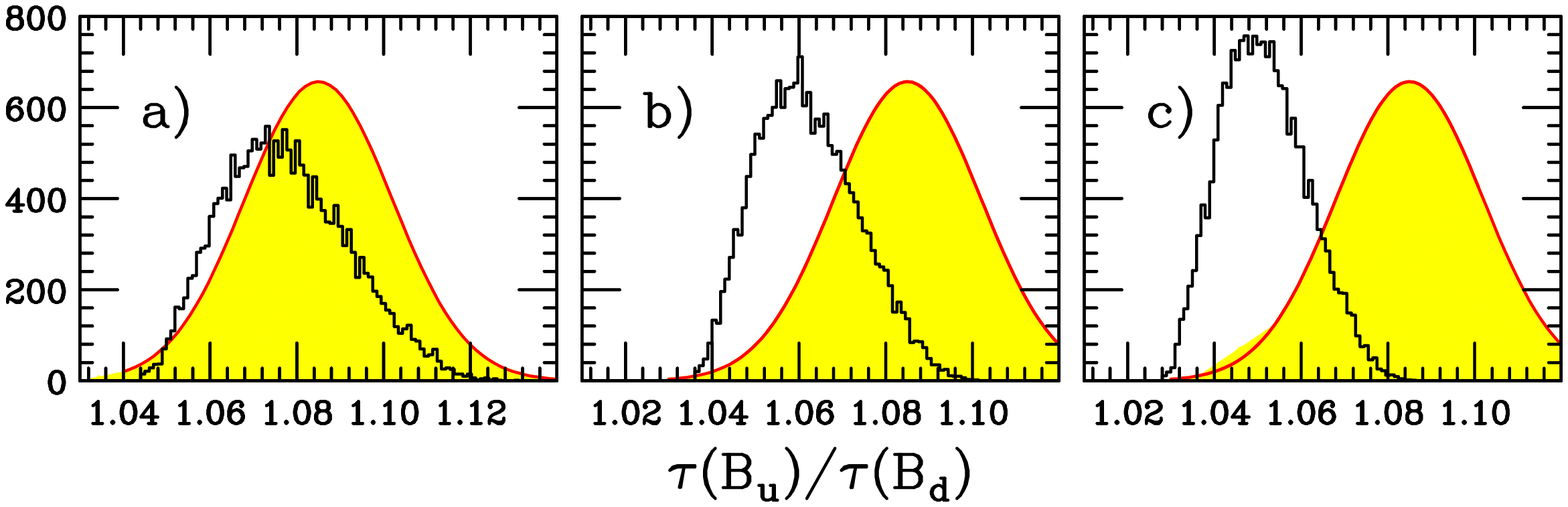, angle=90, width=16cm}}
\centerline{\parbox{14cm}{\caption{\label{fig:randU}
Same as Fig.~\ref{fig:randLam} for $\tau(B_u)/\tau(B_d)$.}}}
\end{figure}

\begin{figure}[!htb]
\centerline{\epsfig{file=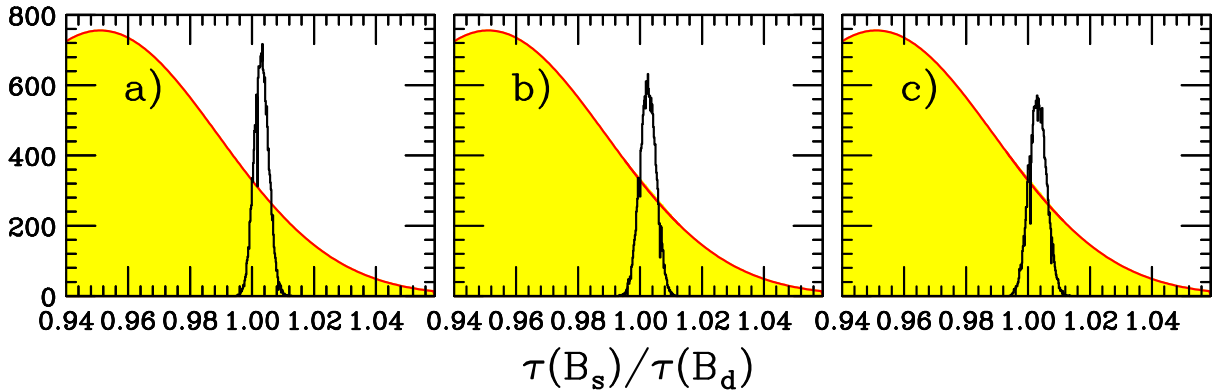, angle=90, width=16cm}}
\centerline{\parbox{14cm}{\caption{\label{fig:randS}
Same as Fig.~\ref{fig:randLam} for $\tau(B_s)/\tau(B_d)$.}}}
\end{figure}

Since the computation of $1/m_b$ corrections to spectator effects was performed
at leading order in QCD, we also studied the scale dependence of the result.
A comparison between the results computed with leading logarithmic accuracy (LL),
next-to-leading log accuracy (NLL) and the total NLL including $1/m_b$ and $1/m_b^2$
corrections is plotted for $\tau(\Lambda_b)/\tau(B_d)$ in Fig.~\ref{fig:mu} for a
renormalization scale parameter interval $m_b/2 \leq \mu \leq 2 m_b$, assuming
``factorization'', {\it i.e.} setting all the $\beta$-parameters in
Eqs.~(\ref{MEMeson}), (\ref{MEBaryon}) to the value
calculated using our ``factorization'' ansatz. Figs.~\ref{fig:mup} and~\ref{fig:mus} show
analogous plots for $\tau(B_u)/\tau(B_d)$ and $\tau(B_s)/\tau(B_d)$. As can be seen from these
graphs, the scale dependence of the predictions is very mild.
%
%
Next, we fix the scale in our calculations and vary the values of
parameters of the matrix elements. We adopt the statistical approach for
presenting our results and generate 20000-point probability distributions of the
ratios of lifetimes obtained by randomly varying our parameters within a
$\pm 30\%$ interval around their ``factorization'' values, for three different
scales $\mu$. The decay constants $f_{B_q}$ and b-quark pole mass $m_b$ are
taken to vary within $1\sigma$ interval indicated above.
The results are presented in Figs.~\ref{fig:randLam},~\ref{fig:randU},
and \ref{fig:randS}. These figures represent the main result of this paper.
They show a nice agreement between the experimental results and
theoretical predictions for all of the ratios of lifetimes of heavy
hadrons. In fact, it is a relatively ``large'' ratio of $\tau(B_s)/\tau(B_d)$
that can pose a potential problem if future measurements at Fermilab and CERN
would find the mean value to stay the same with its error bars shrinking.
Upcoming Fermilab and CERN measurements would again clarify this issue.

It also appears that the theoretical errors in the ratio of $\tau(B_s)/\tau(B_d)$
are much smaller than the other two. We believe that the main reason for it is
the fact that $1/m_b^3$ and higher spectator corrections to both $B_s$ and $B_d$ inclusive
widths are given by the same type of diagrams. In other words,
apart from the different charm-quark mass dependence (which comes at the level of
$m_c^2/m_b^2$ corrections to $1/m_b^3$ effects), both widths of $B_s$ and $B_d$ are
affected by the weak scattering contributions in the same way. Thus, besides possible
$SU(3)$-breaking effects in the matrix elements of the 4-fermion operators, the contributions
of spectator interactions to the numerator and denominator of the ratio of $B_s$ and $B_d$
widths cancel each other to a large degree. This effect persists for the
class of $1/m_b$ corrections to spectator effects discussed here as well.

\section{Conclusions}\label{Conclusions}

We computed subleading $1/m_b$ and $1/m_b^2$ corrections to the spectator effects
driving the difference in the lifetimes of heavy mesons and baryons. Thanks to the same
$16\pi^2$ phase-space enhancement as $1/m_b^3$-suppressed spectator effects, these
corrections constitute the most important set of $1/m_b^4$-suppressed corrections.

We showed that a set of $1/m_b$-corrections to spectators effects can be
parametrized by eight new parameters $\beta_i$ and
$\widetilde{\beta}_i$ ($i=1,...,4$) for heavy mesons and eight new
parameters $\beta_i^\Lambda$ and $\widetilde{\beta}_i^\Lambda$ for
heavy baryons. Although model-independent values of these parameters
will not be known until dedicated lattice simulations are performed,
we present an estimate of these parameters based on quark model
arguments. Compared to \cite{Gabbiani:2003pq}, we kept all matrix
elements, including those vanishing in the factorization
approximation. This is important, as the Wilson coefficients of those
operators are larger than the ones multiplying the operators whose
matrix elements survive in the $N_c \to \infty$ limit. We also perform
studies of convergence of this expansion by computing a set of
$1/m_b^2$-corrections to spectator effects and estimating their size
in factorization. The expansion appears to be well-convergent.

The main result of this paper are Figs.~\ref{fig:randLam},~\ref{fig:randU}, and
\ref{fig:randS}, which represent the effects of subleading spectator effects on
the ratios of lifetimes of heavy mesons and baryons.
We see that subleading corrections to spectator effects affect the ratio of heavy meson
lifetimes only modestly, at the level of a fraction of a percent. On the other hand,
the effect on the $\Lambda_b$-$B_d$ lifetime ratio is quite substantial, at the level of
$-3\%$. This can be explained by the partial cancellation of WS and PI effects in
$\Lambda_b$ baryon and constructive interference of $1/m_b$ corrections to the spectator
effects.

There is no theoretically-consistent way to translate the histograms of
Figs.~\ref{fig:randLam},~\ref{fig:randU}, and \ref{fig:randS} into numerical predictions
for the lifetime ratios. As a useful estimate it is possible to fit the histograms to Gaussian
distributions and extract theoretical predictions for the mean values and deviations of the
ratios of lifetimes. Predictions obtained this way should be treated with care, as it is not
expected that the theoretical predictions are distributed according to the Gaussian
distribution. This being said, we proceed by fitting the distributions to Gaussian and,
correcting for the scale uncertainty [varied in the interval ($m_b/2,2m_b$)] by inflating
the error bars, extract the ratios $\tau(B_u)/\tau(B_d)= 1.06 \pm 0.02$,
$\tau(B_s)/\tau(B_d)= 1.00 \pm 0.01$, and $\tau(\Lambda_b)/\tau(B_d)= 0.86\pm 0.05$.
In these equations the mean values are taken at the scale $\mu=m_b$. This brings the
experimental and theoretical ratios of baryon and meson lifetimes into agreement.

\section*{Acknowledgments}

We would like to thank N.~Uraltsev and M.~Voloshin for helpful discussions and
A.~Lenz for pointing out a numerical mistake in the first version of the manuscript.
A.A.P. thanks the Fermilab Theory Group for hospitality where part of this work was completed.
This work was supported in part by the U.S.\ National Science Foundation
under Grant PHY--0244853, and by the U.S.\ Department of Energy under
Contract DE-FG02-96ER41005.


\appendix

\setcounter{section}{0}
\renewcommand{\thesection} {Appendix \Alph{section}}
\renewcommand{\theequation} {\Alph{section}.\arabic{equation}}
\section{\hspace{-20pt}. Four-quark operators}\label{fqope}
\setcounter{equation}{0}

This paper uses the basis of $1/m_b^3$ operators consistent with Ref.~\cite{Bigi:1994wa},
but different from Refs.~\cite{Neubert:1996we,Ciuchini:2001vx}, given
the relative simplicity of the higher-order operators in our
basis. In the above calculations we used the following notation for
local four-quark operators:
\bea
O_1^q &=& \bar b_i\gamma^{\mu}(1-\gamma_5)b_i\bar q_j\gamma_{\mu}(1-\gamma_5)q_j, ~~
O_2^q = \bar b_i\gamma^{\mu}\gamma_5b_i\bar
q_j\gamma_{\mu}(1-\gamma_5)q_j,\nonumber \\
O_3^q &=& \bar b_i\gamma^{\mu}(1-\gamma_5)b_i\bar
q_j\gamma_{\mu}\gamma_5q_j,
\eea
as well as $\widetilde{O}_i^q$ operators that follow from the expressions for $O_i^q$
by interchanging the color indexes of $b_i$ and $q_j$ Dirac spinors.
Refs.~\cite{Neubert:1996we,Ciuchini:2001vx} use more familiar ``color-singlet'' and
``color-octet'' operators
\bea \label{FourFermion}
Q^q &=& \bar b_L \gamma_\mu q_L \bar q_L \gamma^\mu b_L, \qquad\quad
\ \  Q_S^q = \bar b_R q_L \bar q_L b_R, \nonumber \\
T^q &=& \bar b_L \gamma_\mu t^a q_L \bar q_L \gamma^\mu t^a b_L, \qquad
T_S^q = \bar b_R t^a q_L \bar q_L t^a b_R.
\eea
These operators can be expressed in our basis as
\bea \label{Conv}
Q^q &=& \frac{1}{4} \widetilde{O}^q_1, \qquad\qquad\qquad\quad
Q_S^q = -\frac{1}{8} \widetilde{O}^q_1 -\frac{1}{4} \widetilde{O}^q_2,
\nonumber \\
T^q &=& -\frac{1}{8 N_c} \widetilde{O}^q_1 +\frac{1}{8} O^q_1, \qquad
T_S^q = \frac{1}{8 N_c} \left(\frac{1}{2} \widetilde{O}^q_1 +
\widetilde{O}^q_2\right)-\frac{1}{8}\left(\frac{1}{2} O^q_1 +O^q_2\right) .
\eea
Note that in the predictions for lifetime ratios, large matrix elements
of the ``color-singlet'' operators $Q^q$ and $Q_S^q$ are multiplied by a relatively
small combination of Wilson coefficients, while matrix elements of ``color-octet''
operators come with large Wilson coefficients~\cite{Neubert:1996we,Shifman:wx}.
Nevertheless, the interpretation of matrix elements obtained in the vacuum insertion
approximation (i.e. representing the matrix elements of four-fermion operators
as products of two matrix elements of current operators separated by a vacuum state)
is relatively clear in this basis. For the dimension-6 operators
this procedure introduces four new scale-dependent parameters
$B_i(\mu)$ and $\epsilon_i(\mu)$,
\bea\label{matel}
\frac{1}{2 m_{B_q}} \langle B_q |Q^q| B_q \rangle &=& \frac{f^2_{B_q} m_{B_q}}{8}B_1(\mu), ~~
\frac{1}{2 m_{B_q}} \langle B_q |Q^q_S| B_q \rangle=\frac{f^2_{B_q} m_{B_q}}{8}B_2(\mu),
\nonumber \\
\frac{1}{2 m_{B_q}} \langle B_q |T^q| B_q \rangle &=& \frac{f^2_{B_q} m_{B_q}}{8}\epsilon_1(\mu), ~~
\frac{1}{2 m_{B_q}} \langle B_q |T^q_S| B_q \rangle=\frac{f^2_{B_q} m_{B_q}}{8}\epsilon_2(\mu).
\eea
These parameters can be computed using QCD sum rules, quark models, or on the lattice.
A compilation of various estimates of these parameters can be found in~\cite{Chay:1999pa}.

\section{\hspace{-20pt}. Estimate of matrix elements}\label{matelf}
\setcounter{equation}{0}

We parametrize relevant matrix elements using factorization results as a guiding
principle (a similar approach was used in~\cite{Neubert:1996we,Gabbiani:2003pq,Beneke:1996gn}).
Of course, these matrix elements would eventually have to be computed model-independently,
using lattice QCD or QCD sum rules. In the absence of model-independent
results, we obtain numerical predictions for the ratios of lifetimes,
by relating matrix elements of the computed operators constituting
$1/m_b$ and $1/m_b^2$ effects to the matrix elements of the operators of
dimension 6 ($O_{1,2}$) using quark model. We then use lattice QCD predictions for
the parameters $B_i$ and $\epsilon_i$ and look at the ratio of the subleading-
and leading-order spectator effects. We believe that some errors introduced by our
use of the quark model should partially cancel in these ratios.
The leading-order meson matrix elements are
\bea \label{MEMesonLO}
\langle B_q |O_1^q| B_q \rangle &=& f^2_{B_q} m^2_{B_q} \left(2 \epsilon_1 + \frac{B_1}{N_c}
\right), ~~
\langle B_q |\widetilde{O}_1^q| B_q \rangle = f^2_{B_q} m^2_{B_q} B_1,
\nonumber \\
\langle B_q |O_2^q| B_q \rangle &=& -f^2_{B_q} m^2_{B_q} \left[
\frac{m_{B_q}^2}{\left(m_b+m_q\right)^2} \left(2 \epsilon_2 + \frac{B_2}{N_c}\right)
+ \frac{1}{2}\left(2 \epsilon_1 + \frac{B_1}{N_c}
\right)
\right], \\
\langle B_q |\widetilde{O}_2^q| B_q \rangle &=& -f^2_{B_q} m^2_{B_q} \left[
\frac{m_{B_q}^2}{\left(m_b+m_q\right)^2} B_2 + \frac{1}{2} B_1
\right], \nonumber \\
\langle B_q |O_3^q| B_q \rangle &=& \langle B_q |O_2^q| B_q \rangle, ~~
\langle B_q |\widetilde{O}_3^q| B_q \rangle = \langle B_q |\widetilde{O}_2^q| B_q \rangle.
\nonumber
\eea
The parameters $B_i$ and $\epsilon_i$ represent matrix elements of ``color-singlet'' and
``color-octet'' operators. The above parameters can be computed in QCD sum rules,
quark models, or on the lattice.  Naively, one expects that in the large-$N_c$ limit
$B_i\sim {\cal O}(1), ~\epsilon_i\sim {\cal O}(1/N_c)$.

A parametrization of $1/m_b$-suppressed corrections could, in principle, be quite
simple: after scaling out the proper $m_b$-dependence each new matrix element brings
in a new parameter. This parametrization could be useful for lattice studies of these
matrix elements. Here we shall choose a slightly different
approach. Using the quark model
to {\it guide} our parametrization, we relate our matrix elements to the matrix elements
of singlet and octet operators using Fierz relations. We find
\bea \label{MEMeson}
\langle B_q |R_0^q| B_q \rangle &=& -\frac{1}{2}f^2_{B_q} m^2_{B_q}
\left(2 \widetilde{\beta}_{0} + \frac{\beta_{0}}{N_c}\right)
\left(\frac{m_{B_q}^2}{m_b^2}-1\right),
\nonumber \\
\langle B_q |\widetilde{R}_0^q| B_q \rangle &=&
-\frac{1}{2}f^2_{B_q} m^2_{B_q} \beta_{0}
\left(\frac{m_{B_q}^2}{m_b^2}-1\right),
\nonumber \\
\langle B_q |R_1^q| B_q \rangle &=&
\frac{1}{2} f^2_{B_q} m^2_{B_q}
\left(\frac{\beta_1}{N_c} + 2 \widetilde{\beta_1}\right)
\left(\frac{m_{B_q}^2}{m_b^2}-1\right),
\nonumber \\
\langle B_q |\widetilde{R}_1^q| B_q \rangle &=&
\frac{1}{2} f^2_{B_q} m^2_{B_q} \beta_1 \left(\frac{m_{B_q}^2}{m_b^2}-1\right),
\\
\langle B_q |R_2^q| B_q \rangle &=&
\frac{1}{12} f^2_{B_q} m^2_{B_q}
\left(\frac{\beta_2}{N_c} + 2 \widetilde{\beta_2}\right)
\left(\frac{m_{B_q}^2}{m_b^2}-1\right), \nonumber \\
\langle B_q |\widetilde{R}_2^q| B_q \rangle &=&
\frac{1}{12} f^2_{B_q} m^2_{B_q} \beta_2 \left(\frac{m_{B_q}^2}{m_b^2}-1\right),
\nonumber \\
\langle B_q |R_3^q| B_q \rangle &=&
\frac{1}{12} f^2_{B_q} m^2_{B_q}
\left(\frac{\beta_3}{N_c} + 2 \widetilde{\beta_3}\right)
\left(\frac{m_{B_q}^2}{m_b^2}-1\right), \nonumber \\
\langle B_q |\widetilde{R}_3^q| B_q \rangle &=&
\frac{1}{12} f^2_{B_q} m^2_{B_q} \beta_3 \left(\frac{m_{B_q}^2}{m_b^2}-1\right),
\nonumber
\eea
In the large $N_c$-limit $\beta_i \sim {\cal O}(1)$ and
$\widetilde{\beta}_i \sim {\cal O}(1/N_c)$.
We can use the quark model to relate
$\beta_{0} = \beta_0' (B_2 + B_1/2)$, $\beta_1 = \beta_1' B_1$,
$\beta_{2,3} = \beta'_{2,3} \left[(B_1 - 4 B_2)(1+{m_b^2}/{m_{B_q}^2})+(B_1+2 B_2)\right]$.
The same relations hold for $\widetilde{\beta}_i$ with $B_i \to \epsilon_i$.
Here $\beta_i'$ are used to parameterize the ratios of the matrix elements of
the four fermion operators and $1/m_b$-suppressed operators.
In our numerical studies we took the parameters $B_i,\epsilon_i$ from the lattice calculations
of matrix elements of dimension-6 operators~\cite{Ciuchini:2001vx}
and varied $\beta_i'$ around its quark-model value of unity by $\pm 30\%$.

A similar logic can be used to estimate $1/m_b^2$-suppressed matrix elements. We
simplify the result by using the factorization approximation, i.e. neglecting
all the matrix elements of octet operators, which should suffice for our purposes of studying
the convergence of the series. We obtain
\bea
\langle B_q |W_1^q| B_q \rangle &=& \frac{f^2_{B_q} m^2_{B_q}}{4 N_c}
\left(\frac{m_{B_q}^2}{m_b^2}-1\right)^2, \nonumber \\
\langle B_q |\widetilde{W}_1^q| B_q \rangle &=& \frac{f^2_{B_q} m^2_{B_q}}{4}
\left(\frac{m_{B_q}^2}{m_b^2}-1\right)^2, \nonumber \\
\langle B_q |W_2^q| B_q \rangle &=& -\frac{3 f^2_{B_q} m^2_{B_q}}{8 N_c}
\left(\frac{m_{B_q}^2}{m_b^2}-1\right)^2, \nonumber \\
\langle B_q |\widetilde{W}_2^q| B_q \rangle &=& -\frac{3 f^2_{B_q} m^2_{B_q}}{8}
\left(\frac{m_{B_q}^2}{m_b^2}-1\right)^2,\\
\langle B_q |W_3^q| B_q \rangle &=& -\frac{f^2_{B_q} m_b^2}{8 N_c}
\left(\frac{m_{B_q}^2}{m_b^2}-1\right)^2, \nonumber \\
\langle B_q |\widetilde{W}_3^q| B_q \rangle &=& -\frac{f^2_{B_q} m_b^2}{8}
\left(\frac{m_{B_q}^2}{m_b^2}-1\right)^2, \nonumber \\
\langle B_q |W_4^q| B_q \rangle &=& \langle B_q |W_3^q| B_q \rangle,
~~ \langle B_q |\widetilde{W}_4^q| B_q \rangle = \langle B_q
|\widetilde{W}_3^q| B_q \rangle. \nonumber
\eea
In a similar way, we used the quark-diquark valence-quark model to guide
our parametrizations of baryon matrix elements. The leading order matrix elements
are~\cite{Neubert:1996we,Gabbiani:2003pq}
\bea \label{MEBaryonLO}
\langle \Lambda_b |O_1^q| \Lambda_b \rangle &=&
-\widetilde{B} \langle \Lambda_b |\widetilde{O}_1^q| \Lambda_b \rangle =
\frac{\widetilde{B}}{6}  f^2_{B_q} m_{B_q} m_{\Lambda_b} r, \quad
\nonumber \\
\langle \Lambda_b |O_2^q| \Lambda_b \rangle &=&
-\widetilde{B} \langle \Lambda_b |\widetilde{O}_2^q| \Lambda_b \rangle=
\frac{\widetilde{B}}{6} f^2_{B_q} m_{B_q} m_{\Lambda_b} \delta,
\\
\langle \Lambda_b |O_3^q| \Lambda_b \rangle &=&
-\widetilde{B} \langle \Lambda_b |\widetilde{O}_3^q| \Lambda_b \rangle
= \langle \Lambda_b |O_2^q| \Lambda_b \rangle.
\nonumber
\eea
where $r=\left|\psi^{\Lambda_b}_{bq}(0)\right|^2/\left|\psi^{B_q}_{b\bar
q}(0)\right|^2$ is the ratio of the wave functions at the origin of
the $\Lambda_b$ and $B_q$ mesons, and $\widetilde{B}=1$ in the
valence-quark model. Estimates of $r$ vary from 0.1 to 1.8 and can
potentially be larger~\cite{Neubert:1996we}. Note that $\delta={\cal
O}(1/m_b)$, which follows from the heavy-quark spin symmetry. We
use the same principles to parametrize the $1/m_b$-suppressed
contributions,
\bea \label{MEBaryon}
\langle \Lambda_b |R_0^q| \Lambda_b \rangle &=&
-\frac{{\beta}^{\Lambda}_0}{\widetilde{\beta}^{\Lambda}_0}
\langle \Lambda_b |\widetilde{R}_0^q| \Lambda_b \rangle =
-\frac{{\beta}^{\Lambda}_0}{24}
f^2_{B_q} m_{B_q} m_{\Lambda_b}\left(\frac{m_{\Lambda_b}^2}{m_b^2}-1\right),
\nonumber \\
\langle \Lambda_b |R_1^q| \Lambda_b \rangle &=&
- \frac{{\beta}^{\Lambda}_1}{\widetilde{\beta}^{\Lambda}_1}
 \langle \Lambda_b |\widetilde{R}_1^q| \Lambda_b
\rangle = - \frac{{\beta}^{\Lambda}_1}{24}
f^2_{B_q} m_{B_q} m_{\Lambda_b}
\left(\frac{m_{\Lambda_b}^2}{m_b^2}-1\right),
\\
\langle \Lambda_b |R_2^q| \Lambda_b \rangle &=&
- \frac{{\beta}^{\Lambda}_2}{\widetilde{\beta}^{\Lambda}_2}
 \langle \Lambda_b |\widetilde{R}_2^q| \Lambda_b
\rangle = - \frac{{\beta}^{\Lambda}_2}{48}
f^2_{B_q} m_{B_q} m_{\Lambda_b}
\left(\frac{m_{\Lambda_b}^2}{m_b^2}-1\right),
\nonumber \\
\langle \Lambda_b |R_3^q| \Lambda_b \rangle &=&
- \frac{{\beta}^{\Lambda}_3}{\widetilde{\beta}^{\Lambda}_3}
 \langle \Lambda_b |\widetilde{R}_3^q| \Lambda_b
\rangle = \frac{{\beta}^{\Lambda}_3}{48}
f^2_{B_q} m_{B_q} m_{\Lambda_b}
\left(\frac{m_{\Lambda_b}^2}{m_b^2}-1\right). \nonumber
\eea
Note that the matrix elements contain the parameters
$\beta^{\Lambda}_0$ and $\widetilde{\beta}^{\Lambda}_0$, which are
of the order of $1/m_b$. Thus, these matrix elements only contribute at the
order $1/m_b^2$ corrections to spectator effects (see below). In the
valence-quark approximation these parameters are proportional to $\delta$.
Analogously to the meson case, the quark model relates
$\beta^{\Lambda}_0 = \beta'^{\Lambda}_0 \widetilde{B} \delta$,
$\widetilde{\beta}^{\Lambda}_0 = \widetilde{\beta}^{\Lambda'}_0 \delta$,
$\beta^{\Lambda}_1 = \beta'^{\Lambda}_1 \widetilde{B} r$,
$\widetilde{\beta}^{\Lambda}_1 = \widetilde{\beta}^{\Lambda'}_1 r$,
$\beta^{\Lambda}_{2,3} = \beta^{\Lambda'}_{2,3} \widetilde{B} \left[(r + 4
\,\delta/3) (1+{m_b^2}/{m_{\Lambda_b}^2}) - 2 \,\delta/3 \right]$,
$ \widetilde{\beta}^{\Lambda}_{2,3} =  \widetilde{\beta}^{\Lambda'}_{2,3}\left[(r + 4
\,\delta/3) (1+{m_b^2}/{m_{\Lambda_b}^2}) - 2 \,\delta/3 \right]$.
We use the valence-quark approximation for $1/m_b^2$ corrections. We obtain
\bea
\langle \Lambda_b |W_1^q| \Lambda_b \rangle
&=& -\langle \Lambda_b |\widetilde{W}_1^q| \Lambda_b \rangle =
\frac{f^2_{B_q} m_{B_q} m_{\Lambda_b}}{96}
\left(\frac{m^2_{\Lambda_b}}{m^2_b}-1\right)^2,
\nonumber \\
\langle \Lambda_b |W_2^q| \Lambda_b \rangle
&=& -\langle \Lambda_b |\widetilde{W}_2^q| \Lambda_b \rangle = 0
\\
\langle \Lambda_b |W_3^q| \Lambda_b \rangle
&=& -\langle \Lambda_b |\widetilde{W}_3^q| \Lambda_b \rangle
= \frac{f^2_{B_q} m_{B_q} m_{\Lambda_b}}{192}
\left(\frac{m^2_{\Lambda_b}}{m^2_b}-1\right)^2 \left(1+\frac{m^2_b}{m^2_{\Lambda_b}}\right),
\nonumber \\
\langle \Lambda_b |W_4^q| \Lambda_b \rangle
&=& -\langle \Lambda_b |\widetilde{W}_4^q| \Lambda_b \rangle
= -\langle \Lambda_b |W_3^q| \Lambda_b \rangle, \nonumber
\eea
where we set all the terms of order ${\cal O}(1/m_b^3)$ in the matrix elements
to zero.


\end{document}